# Full-range Gate-controlled Terahertz Phase Modulations with Graphene Metasurfaces

Ziqi Miao[§], Qiong Wu[§], Xin Li[§], Qiong He, Kun Ding, Zhenghua An, Yuanbo Zhang[*] and Lei Zhou[*]

*State Key Laboratory of Surface Physics, Key Laboratory of Micro and Nano Photonic Structures (Ministry of Education), and Department of Physics, Fudan University, Shanghai 200438, China*

[§]These authors contributed equally to this work.

[*] Email: phzhou@fudan.edu.cn; zhyb@fudan.edu.cn.

**Local phase control of electromagnetic wave, the basis of a diverse set of applications such as hologram imaging, polarization and wave-front manipulation, is of fundamental importance in photonic research. However, the bulky, passive phase modulators currently available remain a hurdle for photonic integration. Here we demonstrate full-range active phase modulations in the Tera-Hertz (THz) regime, realized by gate-tuned ultra-thin reflective metasurfaces based on graphene. A one-port resonator model, backed by our full-wave simulations, reveals the underlying mechanism of our extreme phase modulations, and points to general strategies for the design of tunable photonic devices. As a particular example, we demonstrate a gate-tunable THz polarization modulator based on our graphene metasurface. Our findings pave the road towards exciting photonic applications based on active phase manipulations.**

Phase modulation of electromagnetic (EM) waves plays a central role in photonics research. This is best illustrated by Huygens' Principle – the far-field EM wave-front is essentially determined by the phase distribution on a given near-field plane[1]. The ability to control the local EM phase underpins important applications such as holographic imaging[2-5], polarization manipulations[6], and wave-front controls[7-9]. Such phase modulation was conventionally achieved via modulating the refractive index of bulk materials (including both natural materials[10] and recently studied metamaterials (MTM))[11-14]. The dimension of such systems is typically on the order of wavelength[2,10], and therefore too bulky for optical integrations. Metasurfaces, ultra-thin MTMs composed of planar sub-wavelength units, were utilized to achieve phase changes covering full range of $360°$ [8, 9, 15-18]. However, the metasurfaces-based phase modulators currently available are mostly passive elements which cannot be tuned externally.

In this Letter, we experimentally demonstrate full-range gate-tunable THz phase modulation that is realized by an ultra-thin meta-system (thickness $\sim \lambda/10$) integrating graphene and a specially designed metasurface. We show that a gate bias applied on graphene through ion liquid tunes its optical conductivity, turns the coupled system from an under-damped resonator to an over-damped one, and induces drastic modulation on the phase of the reflected wave. We develop an analytical model based on coupled-mode theory[19] (CMT) that captures the essence of our phase-modulation mechanism: a one-port resonator, i.e. a resonator with only reflection channel, is able to drive the phase of the reflected wave across a $\pm 180°$ transition when the losses in the resonator are fine-tuned (in our case by graphene). This is in stark contrast to the two-port resonator (with both reflection and transmission channels) commonly used in previous studies[20], where only a small phase modulation was possible. Our findings represent a significant advance over previous attempts on photonic devices with tunable responses[21-34], and our method points to new design strategies for future active phase modulators. As an example, we present an experimental realization of a tunable THz polarization modulator based on our gate-controlled graphene metasurface.

The structure of our graphene metasurface is shown schematically in Fig. 1a. The metasurface is a five-layer structure that we fabricate sequentially, starting from the bottom layer. An aluminum (Al) film is first evaporated onto a $SiO_2$/Si substrate (not shown in Fig. 1a), and serves as a totally reflective surface for the incident THz wave from above. We then coat a layer of cross-linked photoresist SU8 (MicroChem), followed by an array of Al mesas (100 μm × 80 μm rectangles, shown in Fig. 1b) fabricated on top using standard optical lithography. The Al plane and the mesa (both with a thickness of ~ 50 nm) forms a magnetic resonator as a result of circulating a.c. current induced by normally incident THz wave, which is the basis of our present study.[35, 36] The SU8 layer serves as a spacer between the Al plane and the mesa, whose thickness determines the coupling between the two. Finally, a layer of graphene is transferred onto the structure, and subsequently covered by a layer of gel-like ion liquid. A gate bias, $V_g$, applied between graphene and a side gate tunes the resistance of the graphene[24] (with a peak at the charge neutral Dirac point $V_D$ as shown in Fig. 1c), and modulates the loss of the resonator. We note that the Dirac point of the graphene shifts to $V_D$ ~ 1 V once it is transferred onto the metasurface. The shift is most likely induced by the charged impurities present around graphene, and does not affect our experiment. (Details of the device fabrication and characterization are discussed in the Supplementary Information.) Once the devices are fabricated, we use THz time-domain spectroscopy (THz-TDS) to measure their reflection spectra for both amplitude and phase in a setup schematically depicted in Fig. 1d.

The variation of graphene resistance, induced by the gate, drastically changes the behavior of the resonance in our metasurface, leading to a phase modulation of $\pm 180°$. This is demonstrated on a device with an 85-μm-thick SU8 layer, with the incident THz wave polarized along the long axis of the Al mesas. The reflectance and associated phase spectra for varying gate voltages are shown in the panels **a** and **b**, and panels **c** and **d** of Fig. 2, respectively. Here the relative gate voltage $\Delta V_g$ from the Dirac point, $\Delta V_g = V_g - V_D$, determines the doping level in graphene, and we use the nearly featureless spectrum at the highest doping ($\Delta V_g = 2.02$ V) as our reference. At the Dirac point ($\Delta V_g = 0$) where the graphene is most resistive (therefore has essentially minimal effect on the optical response of the underlying structure), we observe a resonance dip in the reflectance at $f = 0.31$ THz and the phase undergoes a continuous 360° variation across the resonance frequency: all typical intrinsic behaviors of a magnetic resonator[6, 35]. As the doping level starts to increase from zero on the electron side ($\Delta V_g > 0$, Fig. 2a), the reflectance at resonance drops continuously. Meanwhile the resonator maintains an overall magnetic response with the same 360° phase variation, although the bandwidth is narrowed (Fig. 2c). Further increase of the gate bias beyond a critical voltage ($\Delta V_C = 0.76$ V), however, fundamentally changes the behavior of the resonator – the reflectance at resonance starts to increase as $\Delta V_g$ increases (Fig. 2b). More intriguingly, the phase spectrum stops showing a magnetic-resonance character, but rather exhibits a diminishing phase variation centered at 0°. Our results clearly demonstrate an absolute phase modulation of $\pm 180°$ induced by the gate at frequencies around the resonance, and more importantly, our observation indicates a critical transition at $\Delta V_g = \Delta V_C$ in our system. Such a transition is shown more clearly in Fig 4a and 4d, where the critical transition points are seen in the reflectance and phase measured as functions of both gate voltage and frequency. We note that the resonator shows a similar critical behavior on the hole side of the doping ($\Delta V_g < 0$), and a detailed discussion is presented in the Supplementary Information.

The maximum phase modulation in our devices is not limited to $180°$ – in fact, nearly full-range ($360°$) phase tuning can be achieved using two slightly different graphene metasurfaces that are independently gated (elements of device A and B illustrated in Fig. 3a and 3b, respectively). This is made possible by the fact that gate doping modulates the phases from $\sim -180°$ to $0°$ at frequencies right above the resonance in device A (Fig. 3a), and at the same time suppresses the phase from $\sim 180°$ to $0°$ at frequencies just below the resonance in device B (Fig. 3b). As a result, an enormous phase modulation is induced within the frequency interval between the two resonances (shaded region in Fig. 3a and 3b), with a maximum modulation range of $243°$ at 0.48 THz. We note that the range of modulation, although already large, can be further improved by decreasing the overall absorption of graphene and the SU8 spacer (see Supplementary Information). Armed with these results, a plethora of phase-modulation-based applications (such as gate-tunable wave-front control, holographic imaging, and perfect absorber) are now within reach.

How do we interpret the gate-induced critical transition in our graphene metasurfaces – the foundation of our observed large phase modulation? A coherent picture emerges if we treat our system as a one-port resonator within the framework

Page 6 of 20system. Such a transition is shown more clearly in Fig 4a and 4d, where the critical transition points are seen in the reflectance and phase measured as functions of both gate voltage and frequency. We note that the resonator shows a similar critical behavior on the hole side of the doping ($\Delta V_g < 0$), and a detailed discussion is presented in the Supplementary Information.

The maximum phase modulation in our devices is not limited to $180°$ – in fact, nearly full-range ($360°$) phase tuning can be achieved using two slightly different graphene metasurfaces that are independently gated (elements of device A and B illustrated in Fig. 3a and 3b, respectively). This is made possible by the fact that gate doping modulates the phases from $\sim -180°$ to $0°$ at frequencies right above the resonance in device A (Fig. 3a), and at the same time suppresses the phase from $\sim 180°$ to $0°$ at frequencies just below the resonance in device B (Fig. 3b). As a result, an enormous phase modulation is induced within the frequency interval between the two resonances (shaded region in Fig. 3a and 3b), with a maximum modulation range of $243°$ at 0.48 THz. We note that the range of modulation, although already large, can be further improved by decreasing the overall absorption of graphene and the SU8 spacer (see Supplementary Information). Armed with these results, a plethora of phase-modulation-based applications (such as gate-tunable wave-front control, holographic imaging, and perfect absorber) are now within reach.

How do we interpret the gate-induced critical transition in our graphene metasurfaces – the foundation of our observed large phase modulation? A coherent picture emerges if we treat our system as a one-port resonator within the framework of CMT[19]. In what follows, we first establish a generic model for our metasurface and then discuss the crucial role of gate-tunable graphene in governing its behavior.

The resonance formed between the Al mesa and the Al plane in our metasurface can be modeled by a one-port resonator (with resonance frequency $f_0$), driven by an incident wave at frequency $f$. Here the Al plane entirely eliminates the transmission through the resonator, and only the reflection channel needs to be considered (hence the name 'one-port'). The loss in the resonator comes from two sources – absorption within the resonator and radiation to external modes. The former is denoted as the intrinsic loss $\Gamma_i$, and the latter the radiation loss $\Gamma_r$. Analysis based on CMT shows that the reflection coefficient $r$ of such a system can be generally expressed as[19]:

$$r = -1 + \frac{2\Gamma_r}{-i(f-f_0) + \Gamma_i + \Gamma_r}. \tag{1}$$

The evolution of $r$ is best viewed as the Smith curves – traces of $r$ on the complex plane as frequency $f$ increases from $0$ to $\infty$. The curves all start and end at points close to $r = -1$ (Fig. 2e). But they cross the real axis (when in resonance, i.e. $f = f_0$) at locations determined by $r = (\Gamma_r - \Gamma_i)/(\Gamma_r + \Gamma_i)$; the sign of $\Gamma_r - \Gamma_i$ dictates whether the curves encloses the origin. So the relative amplitude of $\Gamma_r$ and $\Gamma_i$ governs the behavior of the reflection phase as $f$ passes the resonance. When $\Gamma_r < \Gamma_i$, the Smith curve does not enclose the origin so that the resulting phase variation is less than $180°$ across the resonance (Fig. 2e, red curve). However, a behavioral transition occurs when $\Gamma_r > \Gamma_i$, and the Smith curve encloses the origin so the reflection phase undergoes a full $360°$ variation (Fig. 2e, blue curve). A transition from an under-damped to an over-damped oscillator takes place at the critical damping ($\Gamma_r = \Gamma_i$) as we increase $\Gamma_i$ (see Fig. 2e). Such a transition fully explains the critical transition that we observed in our metasurfaces (Fig. 2, panels a, b, c and d; Fig. 4a and 4d).

Our resonator model is corroborated by finite-difference-time-domain (FDTD) simulations performed on realistic structures. The simulations quantitatively reproduce all our experimental observations in Fig 2 and 4. They further uncovers the crucial role played by graphene in driving the critical transition: doping of graphene essentially increases the total intrinsic loss $\Gamma_i$ of our device, but has limited influence on the radiation loss $\Gamma_r$, which is determined by the geometry of the metasurface instead (see Supplementary Information for details).

Precious insights on the working mechanism of our gate-tunable metasurface can be gained by further analysis of our FDTD simulations. At low graphene doping ($\Gamma_i < \Gamma_r$), the metasurface (including graphene) is more transparent. Simulations show that THz waves penetrate deeper into the resonator before they are absorbed, and the near-field coupling between the Al mesa and plane forms a magnetic resonance. At high doping levels ($\Gamma_i > \Gamma_r$), however, the absorption is so large that waves cannot reach the Al plane to establish the near-field coupling between the mesa and plane (see Supplementary Information). The system thus behaves as an electric reflector with small phase variation across the resonance. It now becomes clear that the graphene, under the gate-control, serves as a tunable lossy medium; the variable loss breaks the delicate balance between $\Gamma_i$ and $\Gamma_r$, and induces a critical transition in the metasurface accompanied by large phase modulation.

Because $\Gamma_r$, and to a large extent $\Gamma_i$, depend critically on the metasurface geometry, our model further predicts that the critical transition can also be tuned by the structure of the metasurface. Indeed, FDTD simulations show that thinner spacer layer leads to stronger near-field coupling between two metallic layers, resulting in larger Q-factor of the resonance (and therefore a smaller $\Gamma_r$). A smaller $\Gamma_i$ is then needed at the critical transition, so decreasing the spacer layer thickness shifts the transition point towards low doping levels in graphene. Further decrease in the thickness eventually makes $\Gamma_r$ smaller than $\Gamma_i$ even without contribution from graphene. Doping graphene now only makes the metasurface more over-damped since doping can only increase $\Gamma_i$. Critical transition point totally disappears as a result. These predictions are finally compared against the gate-frequency phase diagrams obtained on metasurfaces with 60 μm (Fig. 4b and 4e) and 40 μm (Fig. 4c and 4f) spacer layers, and the excellent quantitative agreements provide additional supports for our model and simulations (see Supplementary Information for details).

Having understood the working mechanism of our metasurface, we are now in a position to formulate a general strategy for designing tunable meta-structures with large phase modulation. We start with an analysis of the nature of the critical transition point – a precondition for the large phase modulation. At the critical point, $\Gamma_r = \Gamma_i$ implies that the system acts as a perfect absorber. For two-port resonators adopted in previous studies[20], the time-reversed version of such a perfect absorber emits symmetrically to both ports of the resonator, which dictates that the critical point (perfect absorption) is reached only when the resonator is excited from two ports simultaneously. In a realistic device where the resonator is excited only from one side, the critical condition can never be satisfied with a finite $\Gamma_i$. The lack of a critical transition is seen clearly in the Smith curves of the transmission coefficient *t* in the two-port model (Fig. 2f). Here all the curves do not enclose the origin, even in the limiting case $\Gamma_i = 0$ (Fig. 2f, broken line), producing limited phase modulation – a problem that dogs previous attempts[20]. The analysis of these competing schemes leads us to a general guideline for designing graphene-based phase modulators: choose systems that support critical point and use graphene as a tunable loss to access both sides of the critical transition. Detailed discussions are presented in Supplementary Information.

Finally, as a demonstration we present a gate-tunable polarizer based on graphene metasurface as shown in Fig. 5e. Here the Al mesas are replaced by stripes, so that the system exhibits magnetic resonance only for the polarization $\vec{E} \parallel \hat{x}$. Upon gating, graphene significantly modulates both the reflectance and the phase spectra (Fig. 5a and 5b), due to the critical transition accompanied by a large phase modulation. On the other hand, the reflection for waves polarized along *y* direction is hardly tuned by graphene (See Supplementary Information) due to the lack of a resonance. Such an anisotropic response can be readily utilized for polarization control. As shown in Fig. 5c and 5d, the reflectance ratio $|r_x|/|r_y|$ and the phase difference $\phi_x - \phi_y$ along *x* and *y* are effectively tuned by the gate, leading to a dramatic modulation on the polarization state of the reflected wave. As a particular example, we fix the frequency at a typical 0.63 THz, and linearly polarize the incident wave at an angle $32°$ with respect to the *x* axis. The reflected wave is now elliptically polarized due to the anisotropic response of our polarizer. The polarization of the reflected wave, characterized by two parameters (namely the ratio between short- and long- axes of the ellipse ($S/L$) and the angle $\theta$ of the long axis, Fig. 5e inset), is now drastically tuned by graphene under gate-control (Fig. 5f).

In conclusion, we demonstrate full-range THz phase modulation tuned by a gate. This is achieved on metasurfaces integrating magnetic resonators with gate-controlled graphene, with overall thickness down to deep sub-wavelength regime. A one-port resonator model is able to capture the essential features of our metasurface. The model further reveals the important role of graphene as a gate tunable loss that modulates the critical transition in the resonator, leading to extreme phase modulation. Our full-range tunable local phase control opens the door to exciting photonic applications in the THz regime. A gate-tunable polarizer is presented as an early demonstration of the capability of our graphene metasurfaces.

## Methods

**THz time-domain measurement**. We use a commercial TDS system (Zomega-3, Zomega Terahertz Corporation) to perform the reflection measurement. The THz source is generated from a GaAs emitter illuminated by a femtosecond fiber laser pulse train with a wavelength of 780 nm, repetition rate of 80 MHz, and pulse width of 125 fs. The horizontal electric field of reflected THz wave is detected via a typical electro-optic sampling method (with ZnTe crystal) in time domain. The THz section of the setup is purged with dry $N_2$ gas to avoid the finger-print absorption of THz wave in air. The spectral amplitude and phase are retrieved from the measured data via numerical Fourier transformation.

**Acknowledgements**

We thank Feng Wang and Meng Qiu for helpful discussions. Part of the sample fabrication was performed at Fudan Nano-fabrication Lab. Z.M., X.L., Q.H., K.D. and L.Z. are supported by National Natural Science Foundation (NSF) of China under grant nos. 11474057 and 11174055, Program of Shanghai Subject Chief Scientist under grant no. 12XD1400700. Q.W. and Y.Z. acknowledge financial support from the National Basic Research Program of China (973 Program) under grant nos. 2011CB921802 and 2013CB921902, and from the NSF of China under grant no. 11034001.

**Author contributions**

Z.M. did the calculations and simulations, and designed the metasurfaces. Q.W. and X.L. fabricated the samples. Z.M., Q.W., X.L. and Q.H. did the THz measurements. K.D. helped with the simulations. Z.A. helped with sample fabrications. L.Z. and Y.Z. initiated, designed and co-supervised the project. Z.M., L.Z. and Y.Z. wrote the manuscript and all authors commented on it.

**Figure captions**

**Figure 1 | Device geometry and experimental setup. a**, Schematic view of the structure of our device. The metasurface consists of an array of Al mesas and a continuous Al film (yellow) separated by a SU8 spacer (blue). CVD-grown monolayer graphene is transferred on top of the metasurface, and subsequently covered by a layer of ion gel. A gate voltage applied between the graphene and a side gate (not shown) controls the doping level in graphene. **b**, Optical image of a typical device viewed from the top. **c**, Gate-dependent resistance of CVD-grown graphene transferred on silicon wafer (blue) and on metasurface (red). **d**, Schematics of our THz-TDS measurement system.

**Figure 2 | Gate-controlled reflection modulation and our resonator model based on CMT. a-b**, Reflectance ($R=|r|^2$) as the relative gate voltage $\Delta V_g$ is varied. A critical transition takes place at $\Delta V_c = 0.76$ V (see text). The reflectances at $\Delta V_g \leq \Delta V_c$ and $\Delta V_g \geq \Delta V_c$ are shown in **a** and **b**, respectively. The spectra are taken on a device with an 85-$\mu m$-thick SU8 spacer. The lateral dimension of the Al mesas is $80\ \mu m \times 100\ \mu m$, and the mesa array has periods of 100 μm and 150 μm in *x* and *y* direction, respectively. The spectrum taken at $\Delta V_g = 2.02$ V is used as the reference. **c** and **d**, Phase modulations for $\Delta V_g \leq \Delta V_c$ and $\Delta V_g \geq \Delta V_c$, respectively. Data are taken on the same device measured in **a** and **b**. **e,** Smith curves of the reflection coefficients *r* in the one-port model illustrated in the inset. The blue, black and red curves represent under-damping, critical damping and over-damping behaviors,

respectively. **f,** Smith curves of the transmission coefficient *t* in the two-port resonator model (inset). No critical transition is present, even in the limiting case $\Gamma_i = 0$ (broken line).

**Figure 3 | Full-range gate-controlled phase modulation.** Gate-dependent reflection phase spectra for device A with a mesa size of 160 μm×110 μm (panel **a**) and device B with a mesa size of 160 μm×100 μm (panel **b**). For both devices, the thickness of the SU8 spacer is 60 μm, and the mesa array has a period of 240 μm in both *x* and *y* directions. The spectrum taken on an Al mirror is used as the reference. Large phase modulation is obtained in the shaded region.

**Figure 4 | Reflectance and phase modulation by metasurfaces with varying spacer thicknesses. a, b** and **c,** Reflectance modulations measured as a function of both gate voltage and frequency for devices with SU8 spacer thicknesses of (**a**) 85 μm, (**b**) 60 μm and (**c**) 40 μm, respectively. Data from these three devices are normalized to spectra measured at $\Delta V_g = 2.02$ V, $\Delta V_g = 2.03$ V and $\Delta V_g = 1.73$ V, respectively. **d, e** and **f**, Reflection phase modulation measured as a function of gate voltage and frequency for the same three devices measured in a, b and c. All three devices has a mesa size of 160 μm×100 μm, and the mesa array has a period of 240 μm in both *x* and *y* directions.

**Figure 5 | A gate-tunable THz polarizer.** Reflectance modulation (panel **a**) and

reflection phase modulation (panel **b**) as a function of gate voltage and frequency measured on the polarizer (shown in **e**). The incident THz wave is polarized perpendicular to the stripes of the polarizer. **c,** Reflection amplitude ratio and **d**, phase difference between polarizations along *x* and *y* measured as functions of frequency at varying gate voltages. Both are effectively modulated by the gate voltage. **e.** Optical image of the polarizer. The stripes have a width of 60 μm, and they are spaced 60 μm apart. **f**, Polarization state (see inset) of the reflected THz wave, characterized by the elliptical axis ratio ($L/S$, red curve) and the rotation angle $\theta$ (blue curve), measured as a function of gate voltage. The working frequency is $0.63\,\text{THz}$. The gate induces drastic modulation of the polarization state.

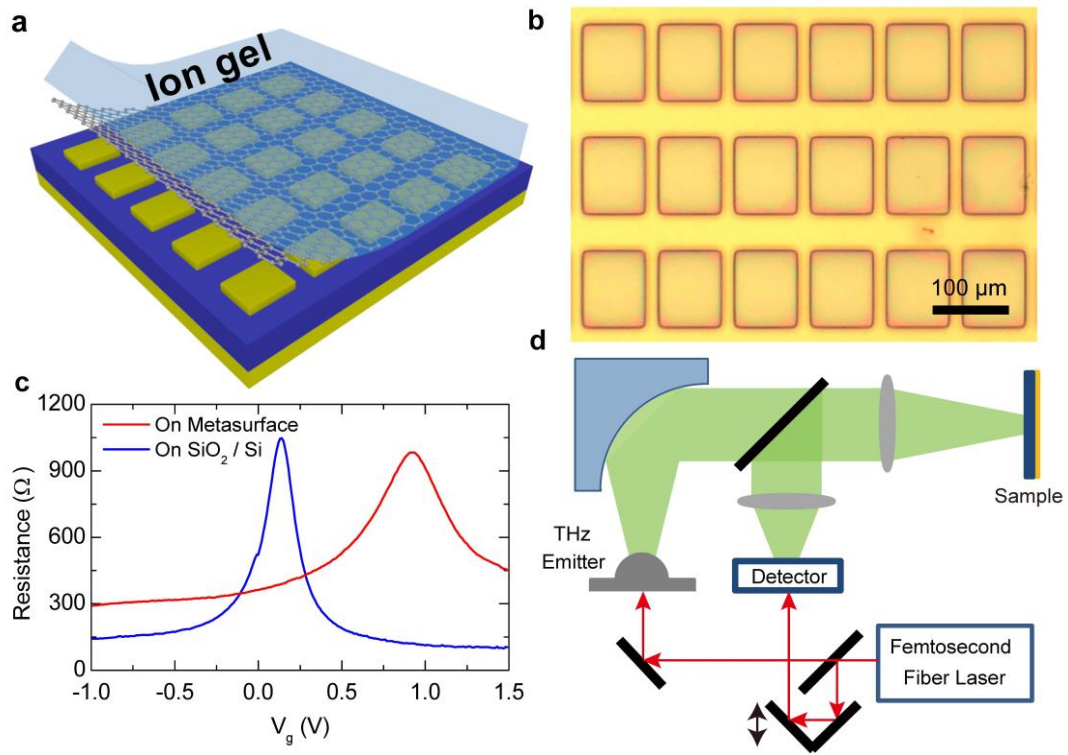

Figure 1, Miao *et al.*

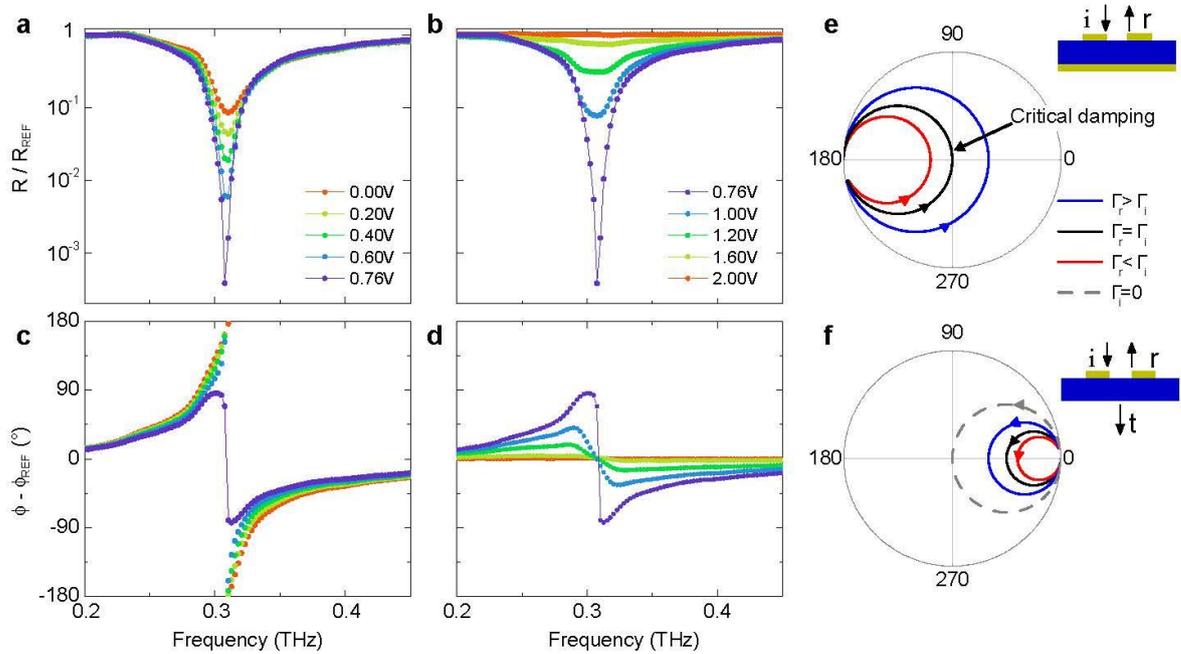

Figure 2, Miao *et al.*

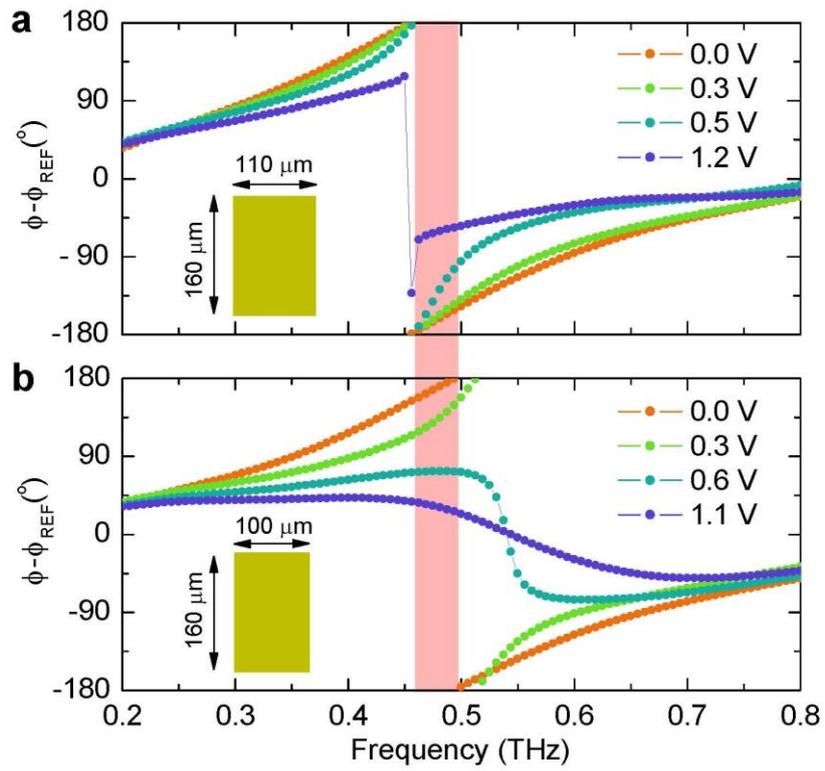

Figure 3, Miao *et al.*

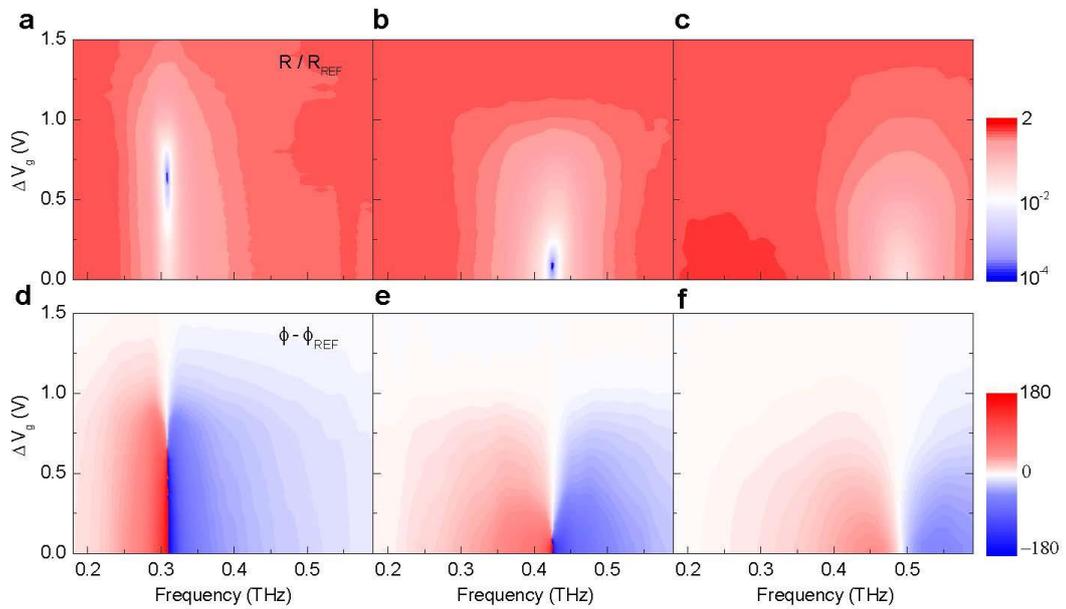

Figure 4, Miao *et al.*

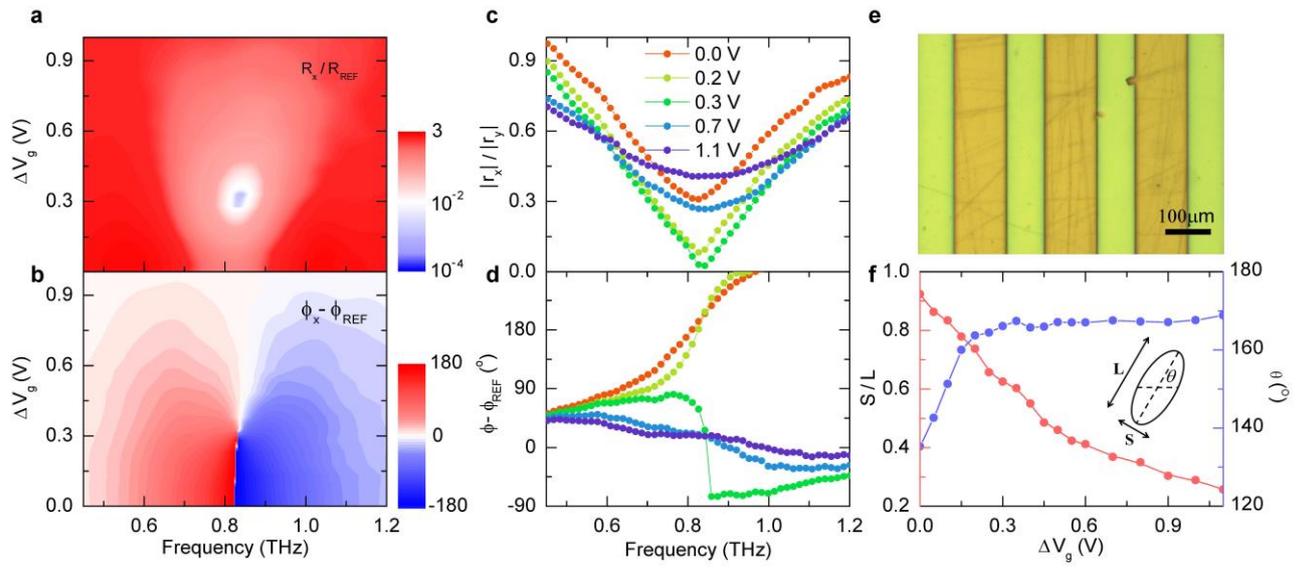

Figure 5, Miao et al.

# Supplementary Information for

# Full-range Gate-controlled Terahertz Phase Modulations with Graphene Metasurfaces

Ziqi Miao, Qiong Wu, Xin Li, Qiong He, Kun Ding, Zhenghua An, Yuanbo Zhang[*] and Lei Zhou[*]

[*] Email: phzhou@fudan.edu.cn; zhyb@fudan.edu.cn.

## I. Preparation of CVD-grown graphene

We grow monolayer graphene on copper foils (Alfa Aesar) using standard chemical vapor deposition (CVD) technique[1,2]. The graphene sample is then transferred onto the target surface following the wet transfer method described in Ref 3, 4. The sample remains uniform after being transferred, and the optical image of a typical graphene sample on $SiO_2$/Si substrate ($SiO_2$ thickness 300 nm) is shown in Fig. S1 inset. Raman spectroscopy measurement (Fig. S1) confirms that the graphene sample is monolayer.

A gate voltage applied between graphene and a side gate modulates the doping level of our graphene sample, with ion-gel serving as the gate medium[5]. The ion gel was made by dissolving 1-ethyl-3-methylimidazolium bis(trifluoromethylsulfonyl)-imide ([EMIM][TFSI]) in poly(styrene-ethylene oxide-b-styrene) (PS-PEO-PS) polymer matrix. Gate voltages up to $\pm 3.0 V$ can be applied without damaging the graphene samples, and the corresponding maximum doping level could reach about $6 \times 10^{13} cm^{-2}$ (Ref. 5).

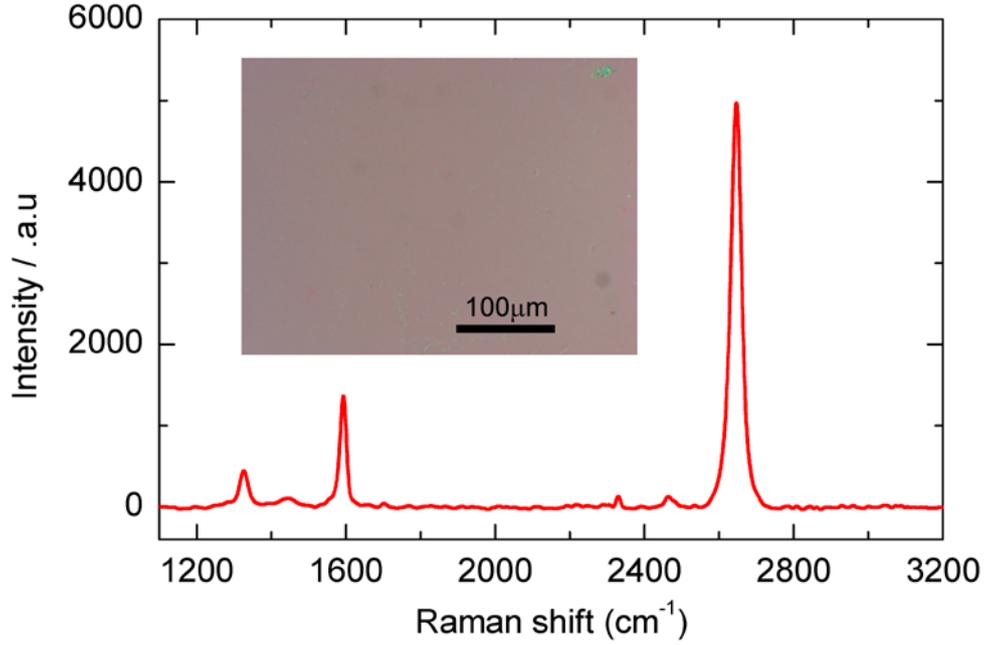

**Fig. S1 | CVD-grown monolayer graphene.** Raman spectrum of our CVD-gown monolayer graphene transferred onto Si substrate covered with 300 nm $SiO_2$. Inset: optical image of a typical monolayer graphene sample on $SiO_2$/Si substrate.

## II. Critical transition modulated by hole-doped graphene

The critical transition in our metasurfaces discussed in the main text can also be modulated by graphene on the hole side of the doping. Here we present data obtained from the same devices studied in Figs. 2 and 4, now with the graphene doped with holes instead of electrons. We find that the metasurfaces exhibit similar behavior as the thickness of the SU8 spacer layer is varied. Specifically, metasurfaces with 85-μm-thick spacer layer (Fig. S2) and 60-μm-thick spacer layer (Fig. S3) feature gate-controlled critical transitions, but the critical transition is not accessible for the metasurface with a 40-μm-thick spacer layer (Fig. S4). We note that the critical gate voltages relative to the charge neutral Dirac point are not strictly the same for electron and hole doping. This is possibly due to the fact that the electron and hole sides of graphene band structure are not exactly symmetric, especially at high doping levels.

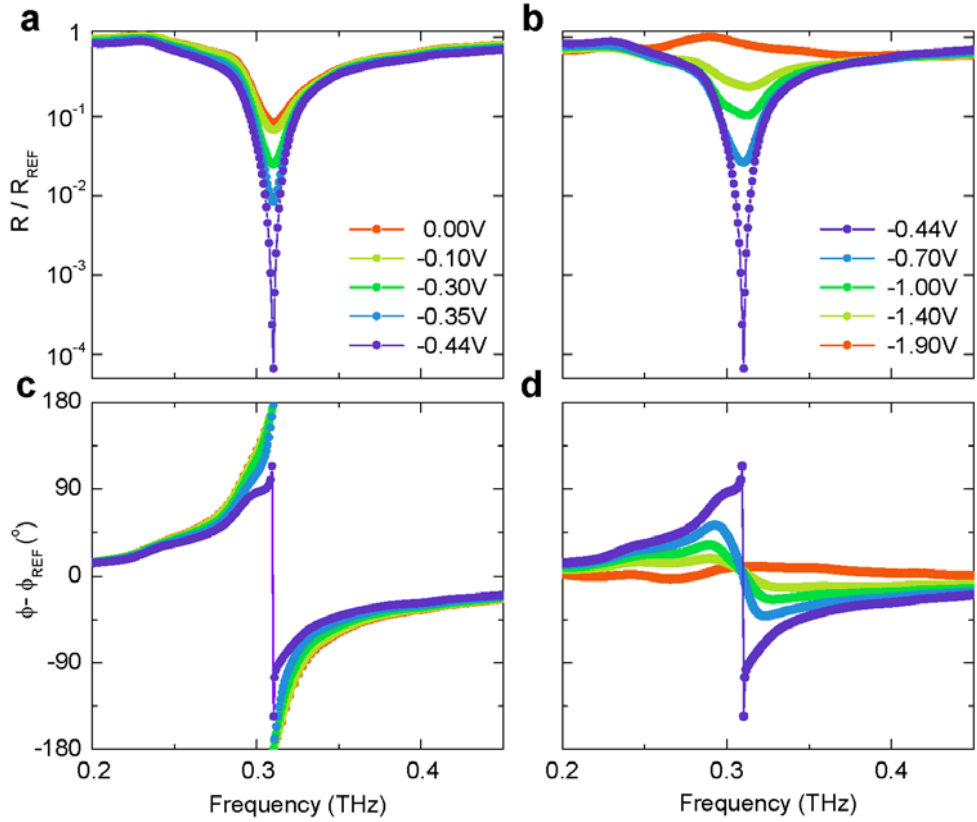

**Figure S2 | Gate-controlled critical transition for device with 85-μm-thick spacer layer under hole-doping. a** and **b**, Reflectance modulation for relative gate-voltage $|\Delta V_g|$ smaller than the critical value $|\Delta V_c| = 0.44$ V and for $|\Delta V_g|$ greater than $|\Delta V_c|$, respectively. **c** and **d**, Phase modulation for $|\Delta V_g| \lesssim |\Delta V_c|$ and $|\Delta V_g| \gtrsim |\Delta V_c|$, respectively. Data are obtained on the same device presented in Fig. 2, Fig. 4(a) and Fig. 4(d).

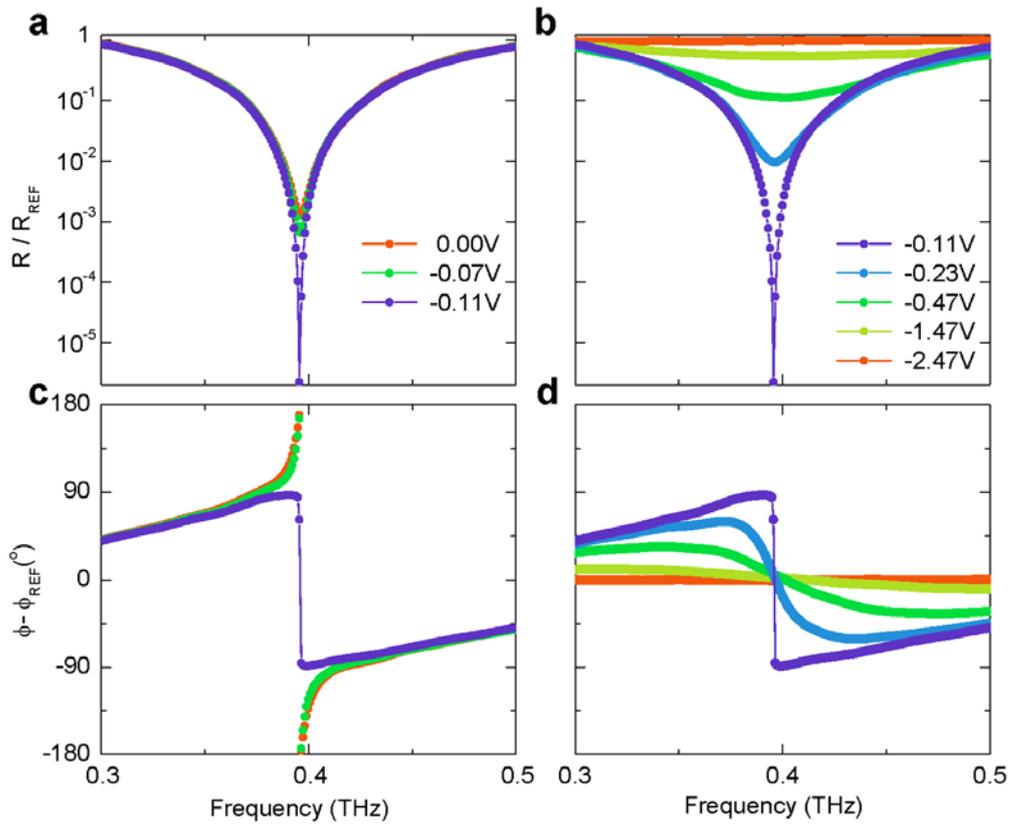

**Figure S3 | Gate-controlled critical transition for device with 60-μm-thick spacer layer under hole-doping. a** and **b**, Reflectance modulation for relative gate-voltage $|\Delta V_g|$ smaller than the critical value $|\Delta V_c| = 0.11$ V and for $|\Delta V_g|$ greater than $|\Delta V_c|$, respectively. **c** and **d**, Phase modulation for $|\Delta V_g| \lesssim |\Delta V_c|$ and $|\Delta V_g| \gtrsim |\Delta V_c|$, respectively. Data are obtained on the same device presented in Fig. 4(b) and 4(e).

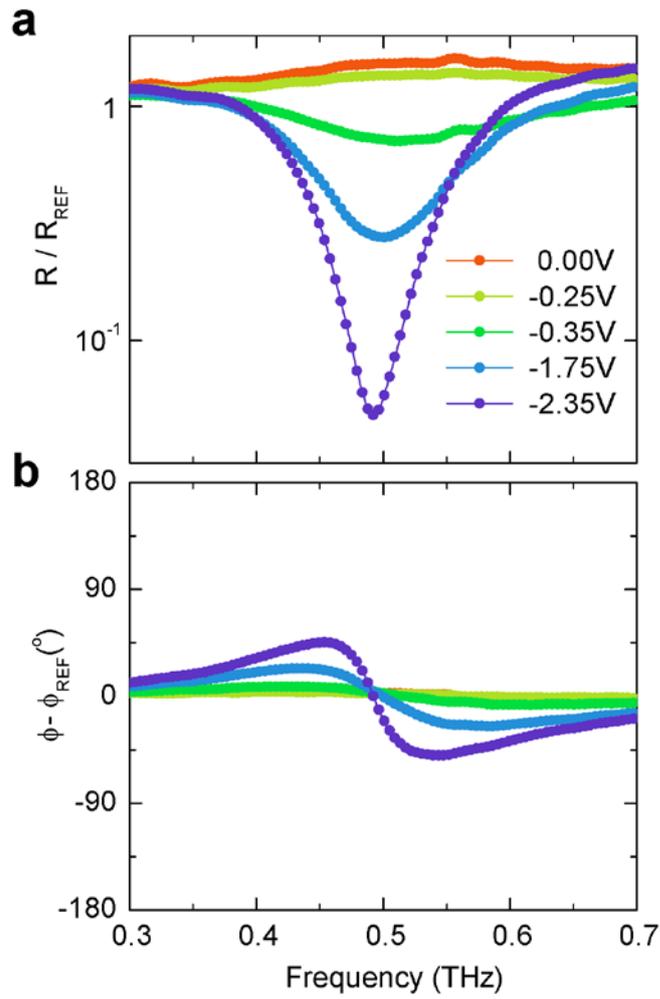

**Figure S4 | Gate-controlled reflection modulation for device with 40-μm-thick spacer layer under hole-doping. a** and **b**, Reflectance and phase modulation, respectively. No critical transition is observed. Data are obtained on the same device presented in Fig. 4(c) and 4(f).

## III. Reflectance and phase of our THz polarizer in *y* direction

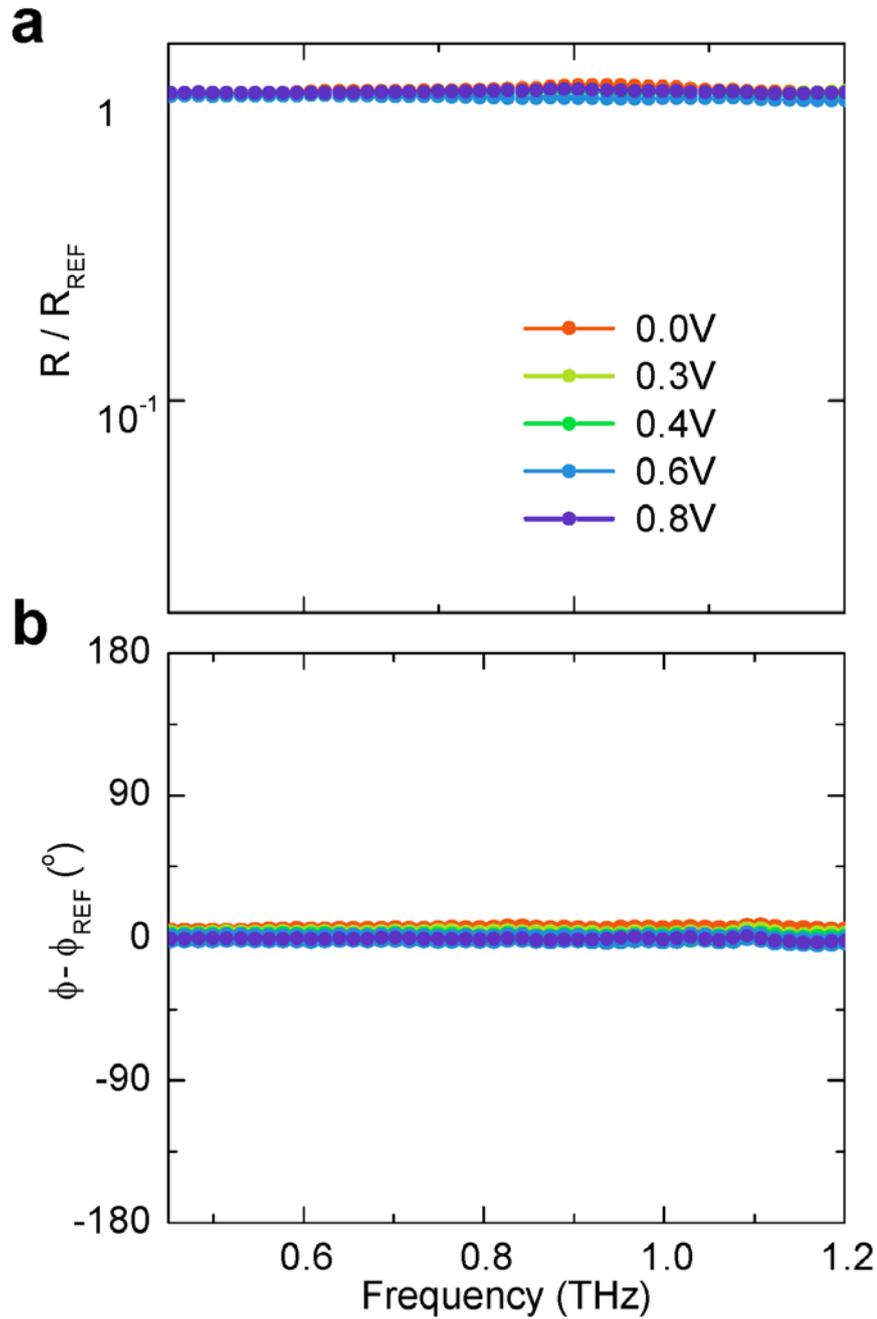

**Figure S5 | (Lack of) reflection modulation by our THz polarizer with incident wave polarized in *y* direction. a,** Reflectance and **b**, phase of the reflected wave as the gate voltage is varied. Incident wave is polarized in *y* direction (along the stripes of the polarizer). Spectrum taken at $\Delta V_g = 1.1$ V is used as the reference.

## IV. Performance of our devices referenced to Al mirror

In the main text, all spectra are referenced against the reference spectra measured under highest doping. While this is enough to illustrate the phase modulation effect (which does not need an absolute reference), one cannot retrieve the working efficiencies of our devices. Here we provide additional data on the absolute reflectance of our devices (referenced to Al mirror). Figure S6 shows the absolute reflectance spectra of the two devices (device A and B) studied in Fig. 3. We also measured the absolute reflectance spectra of the tunable polarizer (the same device studied in Fig. 5), and then retrieved the working efficiency (i.e., the power ratio between the reflected and incident waves) of our device under the experiment condition described in Fig. 5(f). The working efficiency of our device at 0.63 THz is found to be ~ 40% within the range of ours applied gate voltage (Fig. S7).

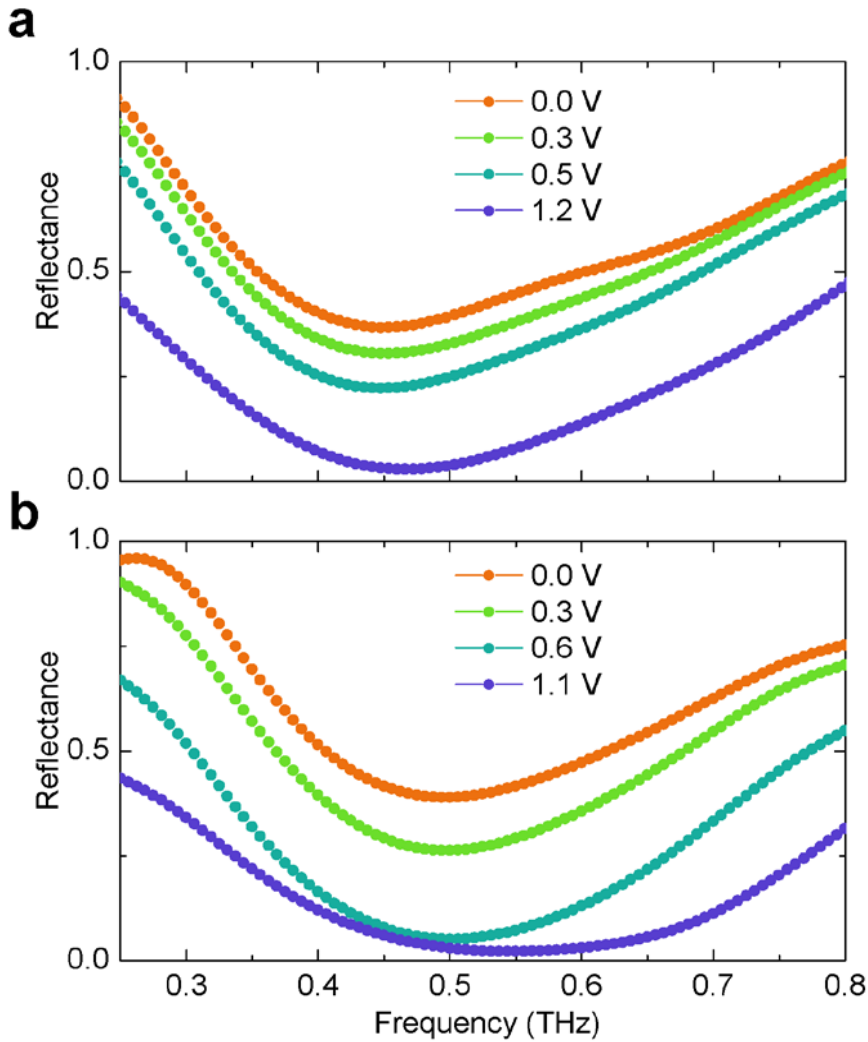

**Figure S6 | Reflectance spectra referenced to Al mirror. a** and **b,** Reflectance spectra of device A and B (the same devices studied in Fig. 3), respectively, as the gate voltage is varied.

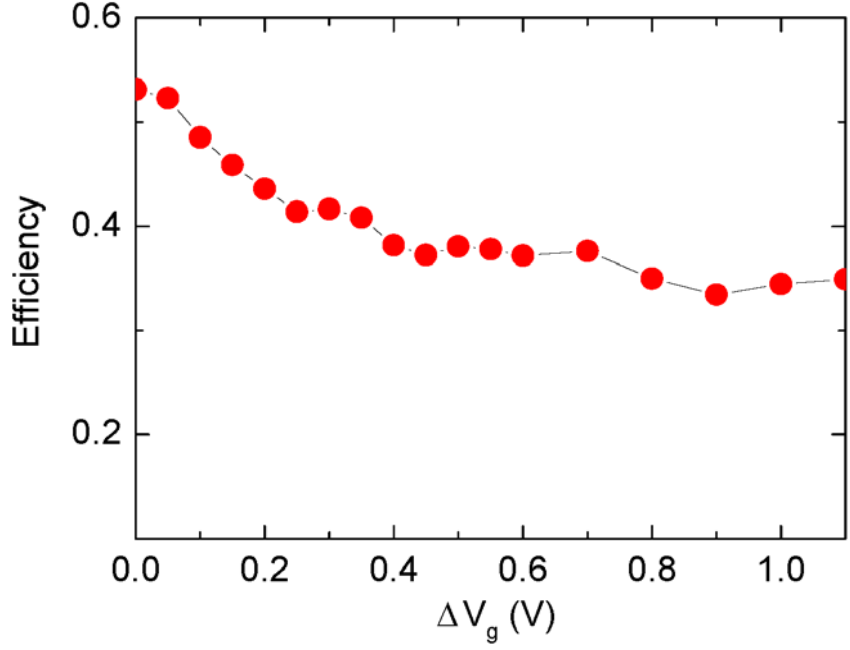

**Figure S7 | Working efficiency of the tunable polarizer.** The efficiency is measured as a function of relative gate voltage, and the data are obtained from the same polarizer studied in Fig. 5 of the main text.

## V. FDTD Simulation of our metasurfaces

### A. Simulation model

The conductivity of graphene in the THz regime can be described by the linear-response theory under random phase approximation that only includes intra-band contributions[6]:

$$\sigma(\omega) \approx \frac{ie^2 v_f \sqrt{n}}{\hbar\sqrt{\pi}(\omega+i\Gamma)}, \quad (S1)$$

where $v_f$ is the Fermi velocity of electrons in graphene ($\sim 9\times 10^5 m/s$)[7], $n$ is the two-dimensional (2D) electron density of graphene and $\Gamma$ is a phenomenological constant accounting for the electron scattering rate (damping). In our FDTD simulations, graphene is modeled as a thin dielectric layer with thickness $d$ that exhibits an anisotropic dielectric function[8] $\vec{\varepsilon}_r = \text{diag}[\varepsilon_r^\parallel, \varepsilon_r^\parallel, 1]$, where

$$\varepsilon_r^\parallel = \sqrt{\varepsilon_{r1}\varepsilon_{r2}} - \frac{\sigma(\omega)}{i\varepsilon_0 \omega d} \quad (S2)$$

takes a standard Drude form. Here $\varepsilon_{r1}$ and $\varepsilon_{r2}$ are dielectric constants of two media surrounding graphene, and $d$ is set as 1 nm. We note that $d$ here is much larger than the real thickness of graphene, 0.3 nm. But it was proved in Ref. 8 that exact value of $d$ does not influence our final result as long as it is much smaller than the wavelength, which is indeed the case in our simulations. We assume that the carrier density $n$ in graphene is related to the gate voltage $\Delta V_g$ through $n = (n_0^2 + \alpha |\Delta V_g|^2)^{1/2}$, with $\alpha$ being the gate capacitance of the ion-gel-based gating scheme and $n_0$ the residue carrier density[9], determined by fitting experimental data.

We treat Al as lossy metal in our FDTD simulations. Best simulation results are obtained when the electric conductivity of Al is set as $10^6 S/m$, which is much smaller than the bulk value $3.5 \times 10^7 S/m$ (Ref. 10). Such a difference between bulk and thin-film Al has also been observed by previous THz measurements[11]. The SU8 spacer is treated as a dielectric insulator with $\text{Re}(\varepsilon) = 3.5$ and $\text{Im}(\varepsilon) = 0.28$. All these parameters were determined through carefully comparing the FDTD results with experimental data on various samples (metasurfaces without graphene).

## B. Simulation results and discussions

We performed extensive FDTD simulations to study all the cases experimentally characterized in the main text. Figures S8 and S9 show the FDTD results corresponding to the cases presented in Fig. 2 and Fig. 4 in the main text, respectively. In our simulations, we set $n_0 = 9.04 \times 10^{11} cm^{-2}$, $\alpha = 3.35 \times 10^{12} cm^{-2} V^{-1}$ for the 85μm case, $n_0 = 2.34 \times 10^{12} cm^{-2}$, $\alpha = 5.89 \times 10^{12} cm^{-2} V^{-1}$ for the 60μm case and $n_0 = 2.31 \times 10^{12} cm^{-2}, \alpha = 6.50 \times 10^{11} cm^{-2} V^{-1}$ for the 40μm case, respectively. We note that while these parameters are of the same order, their absolute values are different, which we attribute to different local environments experienced by graphene in different samples. Our simulations reproduce all essential features, especially the critical transitions tuned by gate and SU8 spacer thickness, observed in our experiments.

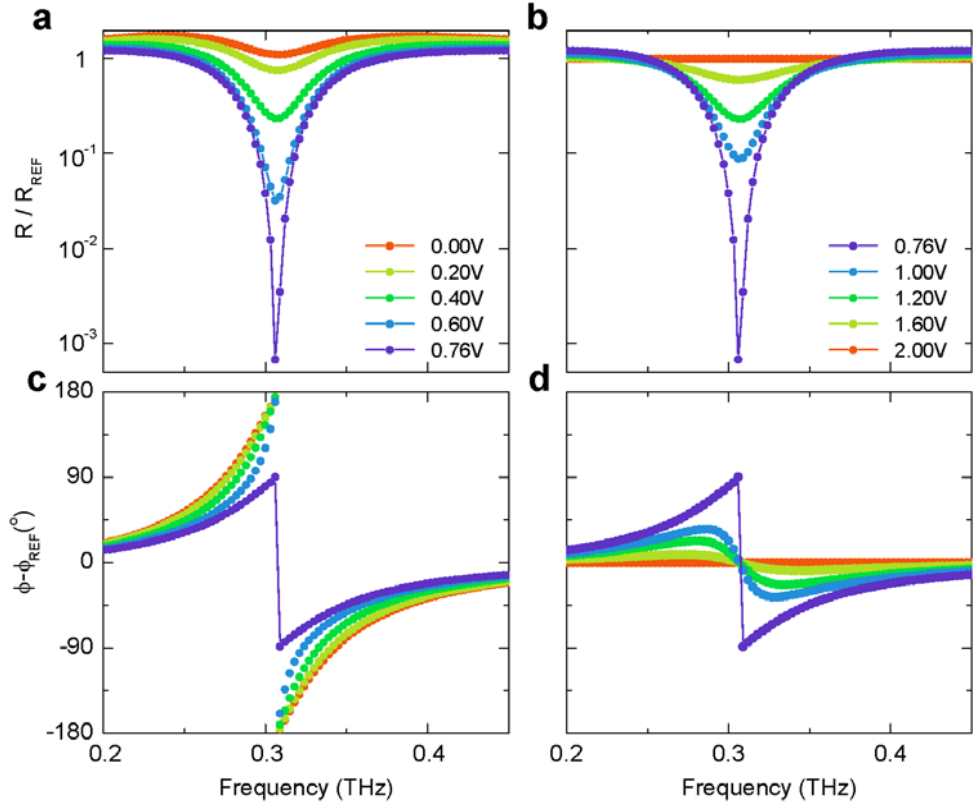

**Figure S8 | FDFD simulations for the experiments described in Fig. 2. a-b**, Reflectance ($R = |r|^2$) as the relative gate voltage $\Delta V_g$ is varied. A critical transition takes place at $\Delta V_c = 0.76$ V. The reflectances for the cases of $\Delta V_g \leq \Delta V_c$ and $\Delta V_g \geq \Delta V_c$ are shown in **a** and **b**, respectively. The spectra are obtained on a device with an 85-μm-thick SU8 spacer. The lateral dimension of the Al mesas is 80 μm × 100 μm, and the mesa array has periods of 100 μm and 150 μm in *x* and *y* direction, respectively. The spectrum taken at $\Delta V_g = 2.02$ V is used as the reference. **c** and **d**, Phase modulations for $\Delta V_g \leq \Delta V_c$ and $\Delta V_g \geq \Delta V_c$, respectively.

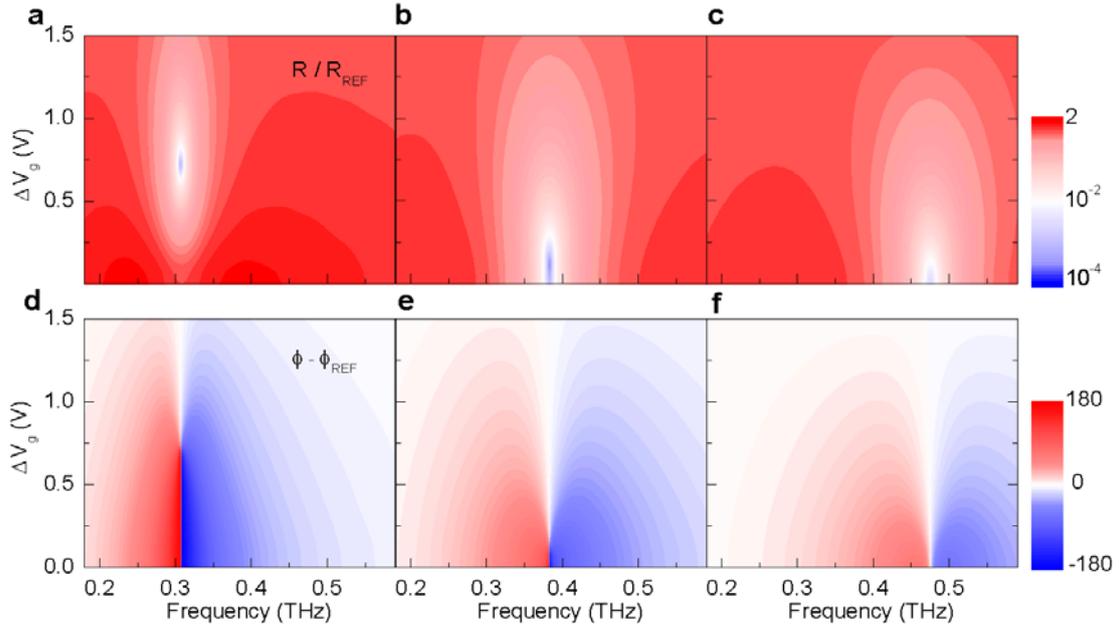

**Figure S9 | FDTD simulations for the experiments described in Fig. 4. a, b** and **c,** Reflectance modulations calculated as a function of both gate voltage and frequency for devices with SU8 spacer thicknesses of $85\,\mu m$, $60\,\mu m$ and $40\,\mu m$, respectively. Data from these three devices are normalized to spectra calculated at $\Delta V_g = 2.02\text{ V}$, $\Delta V_g = 2.03\text{ V}$ and $\Delta V_g = 1.73\text{ V}$, respectively. **d, e** and **f,** Reflection phase modulation calculated as a function of gate voltage and frequency for the same three devices measured in a, b and c. All three devices has a mesa size of $160\,\mu m \times 100\,\mu m$, and the mesa array has a period of $240\,\mu m$ in both *x* and *y* directions.

   To gain a deeper understanding on the nature of the resonance in the under-damped and over-damped regimes, we employed FDTD simulations to calculate the field patterns for two representative cases. The field patterns in the under-damped and over-damped metasurfaces are shown in Fig. S10(a) and Fig. S10(b) respectively. In the case of the under-damped resonator (Fig. S10(a)), waves penetrate deep inside the metasurface to establish the near-field coupling between the two metallic layers, and establish a magnetic resonance. In contrast, waves are directly reflected back by the top mesa layer before they could enter the metasurface (Fig. S11(b)). This picture naturally explains why our metasurface, when located in the over-damped region, behaves as an electric reflector because essentially only the top mesa layer is working to reflect THz waves.

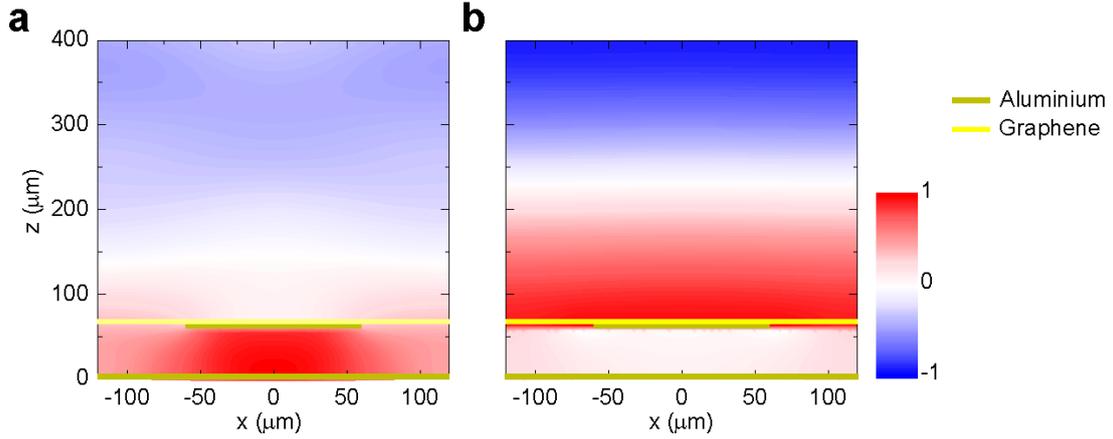

**Figure S10 | Comparison of the field patterns in under-damped and over-damped resonators. a,** Simulated $H_y$ distribution in the $x$-$z$ plane for our graphene metasurface with a 60-$\mu m$-thick SU8 spacer layer. The size of the Al mesa is $160\ \mu m \times 120\ \mu m$, and the mesa array periodicity is fixed at $240\ \mu m \times 240\ \mu m$ (only one mesa is shown here). Carrier density in graphene is set at $2.0 \times 10^{11} cm^{-2}$, which puts the resonator in the under-damped region. **b,** Simulated field distribution when the resonator is in the over-damped region (graphene carrier density $1.45 \times 10^{13} cm^{-2}$). In both cases, the devices are illuminated by normally incident plane waves polarized with $\vec{E} \parallel \vec{x}$

## VI. Graphene as a tunable loss in the metasurface – an analysis based on CMT

To understand the crucial role of graphene in modulating the resonance behavior of our metasurfaces, we analyze the effect of graphene doping based on Eq. (1). We extract the radiative loss and intrinsic loss $\Gamma_r$ and $\Gamma_i$ at different gate voltages by fitting the corresponding FDTD simulation results with Eq. (1). We chose to fit the FDTD simulation results instead of experimental data because a) all experimental features are well reproduced by FDTD simulations, and b) the fitting depends very sensitively on the shape of the curve at the resonance; fluctuations in the measured spectra make fittings of experimental data unreliable. It should be noted that, strictly speaking, the CMT is only valid at frequencies at the resonance. In order to obtain unambiguous fitting results, we perform the fitting procedure in a frequency interval centered at the resonance. We then vary the bandwidth of the interval, and make sure that the obtained fitting results converge and are nearly independent of the bandwidth.

Figure S11 shows how the extracted $\Gamma_r$ and $\Gamma_i$, scaled by the corresponding resonance frequency $f_0$, vary with the gate voltage $\Delta V_g$ in different devices. Two conclusions can be drawn from this figure. First, increasing the carrier density in graphene (through increasing $\Delta V_g$) increases the intrinsic loss $\Gamma_i$, but leaves the radiation loss $\Gamma_r$ almost intact. Second, when the graphene doping level is fixed, metasurface with thinner spacer layer exhibits smaller radiation loss. Therefore, while sections of the resonances in the two metasurfaces with 85μm and 60μm spacer layers locate in the under-damped region, all resonances in the metasurface with a 40μm-thick space layer locates in the over-damped region due to its diminished radiation loss. The latter does not show a transition as $\Delta V_g$ is increased because gating can only increase the intrinsic loss which drives the system even further away from the transition boundary.

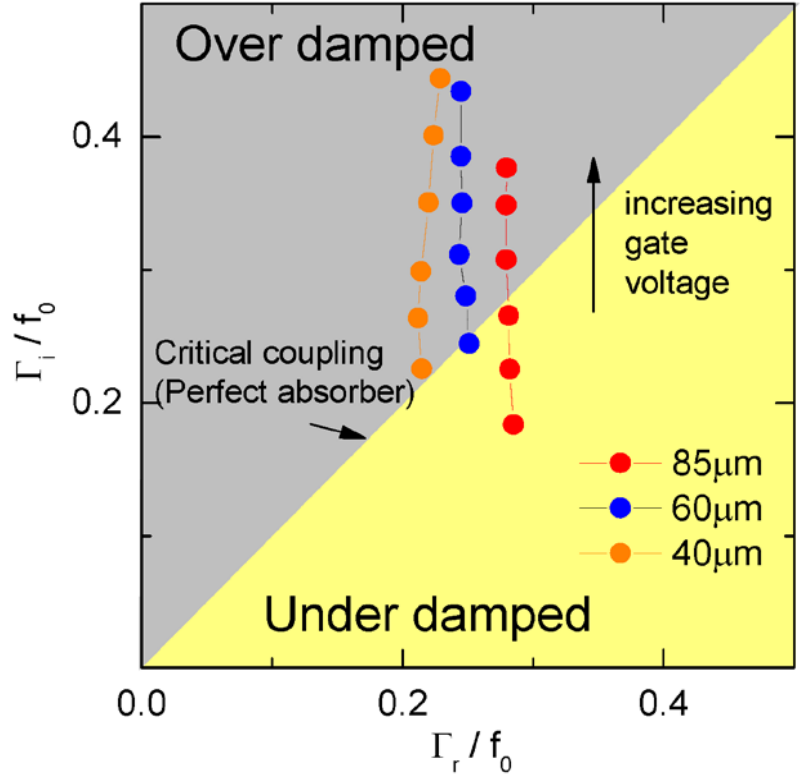

**Figure S11 | Critical transition modulated by doping and spacer layer thickness.** $\Gamma_r$ and $\Gamma_i$ are extracted by fitting FDTD simulations results with Eq. (1). Three device structures (with spacer layer thickness 40 μm, 60 μm and 85 μm) are examined. Gate voltages are varied from $\Delta V_g = 0$ V to $\Delta V_g = 1.5$ V.

It is already clear from Fig. S11 that different spacer layer thickness results in different radiation loss $\Gamma_r$. Here we employ FDTD simulations to further quantify the effect of the spacer layer thickness on $\Gamma_r$. We compute the Q factor (which is inversely propotional to $\Gamma_r$) of a series of devices with varying spacer thickness. To spcifically delineate the effect of spacer thickness on the radiation loss, all materials (both metal and SU8) are assumed dissipationless (i.e. intrinsic loss is set to 0). The Q factor as a function of spacer layer thickness is shown in Fig. S12. A thinner spacer layer leads to an enhanced Q factor, which explains why the radiation loss decreases with reduced spacer layer thickness (Fig. S11). Indeed, metasurface with a thinner spacer has a stronger near-field coupling between two metallic layers, leading to an enhanced Q factor, and therefore a smaller $\Gamma_r$. Such an intriguing effect has already been discussed in a similar system[12].

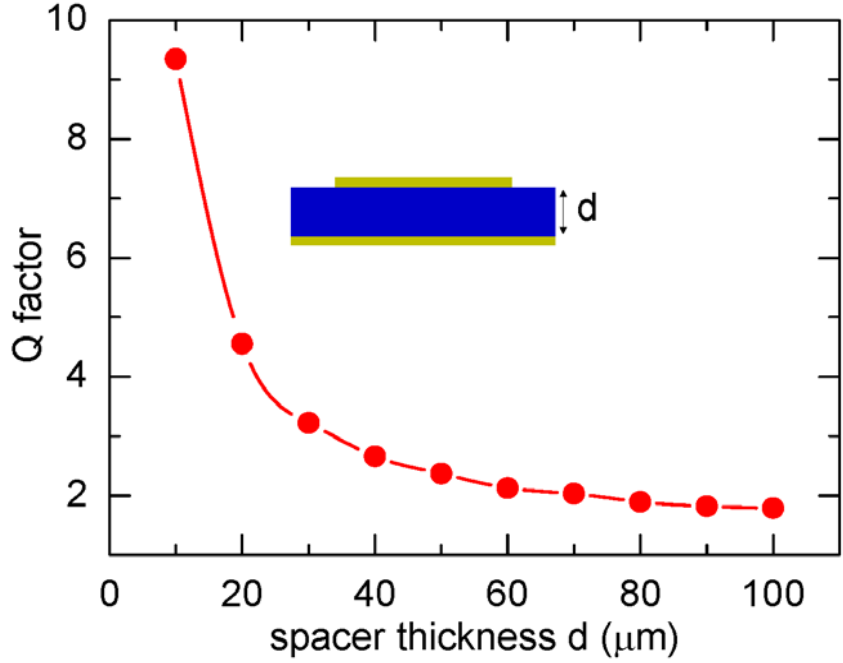

**Figure S12 | Q factors computed as a function of spacer layer thickness.** All materials in the metasurface are assumed dissipationless. The Al mesa layer has exactly the same geometry as those studied in Fig. 4, and only the spacer layer thickness *d* is varied.

## VII. CMT analysis of a two-port resonator

According to the CMT[13], assuming that the background medium (medium without the resonant structure) is perfectly transparent (i.e., $t_0 = 1, r_0 = 0$), we have for the two-port, single-mode model:

$$\begin{cases} r = \dfrac{\Gamma_e}{i(f - f_0) - \Gamma_i - \Gamma_e} \\ t = 1 + \dfrac{\Gamma_e}{i(f - f_0) - \Gamma_i - \Gamma_e} \end{cases}. \quad (S3)$$

The Smith curves for the transmission coefficient *t* have already been plotted in Fig. 2(f) in the main text. Here we present the Smith curves for reflection coefficient *r* in Fig. S13. Again, no critical transition is observed in *r* however the doping level is varied.

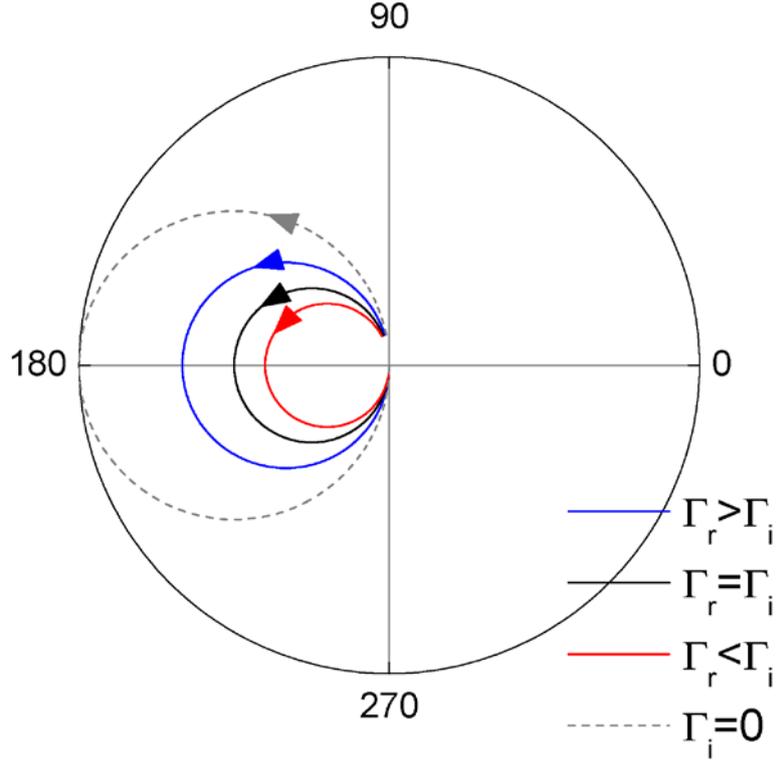

**Figure S13 | CMT analysis of a two-port, single-mode resonator.** Smith curves for $r$ according to Eq. (S3). Four representative cases are plotted: a) $\Gamma_i = 0.05 f_0, \Gamma_r = 0.1 f_0$ (blue curve); b) $\Gamma_i = 0.1 f_0, \Gamma_r = 0.1 f_0$ (black curve); c) $\Gamma_i = 0.15 f_0, \Gamma_r = 0.1 f_0$ (red curve); d) $\Gamma_i = 0, \Gamma_r = 0.1 f_0$ (gray curve).

In general, the background medium may not be perfectly transparent but rather exhibits finite transmission ($t_0$) and reflection ($r_0$). A generalized model based on CMT yields:

$$\begin{cases} r = r_0 + \dfrac{\Gamma_e e^{i\Phi}}{i(f - f_0) - \Gamma_i - \Gamma_e} \\ t = t_0 + \dfrac{\Gamma_e e^{i\Phi}}{i(f - f_0) - \Gamma_i - \Gamma_e} \end{cases} \quad (S4)$$

with $e^{i\Phi} = -(t_0 + r_0)/|t_0 + r_0|$ (Ref. 13). Calculations based on Eq. (S4) show that the Smith curves are now tilted (Fig. S14). But our conclusion (i.e. lack of critical transition in two-port resonators) still stands, even in the limiting case $\Gamma_i = 0$ (broken lines, Fig. S14).

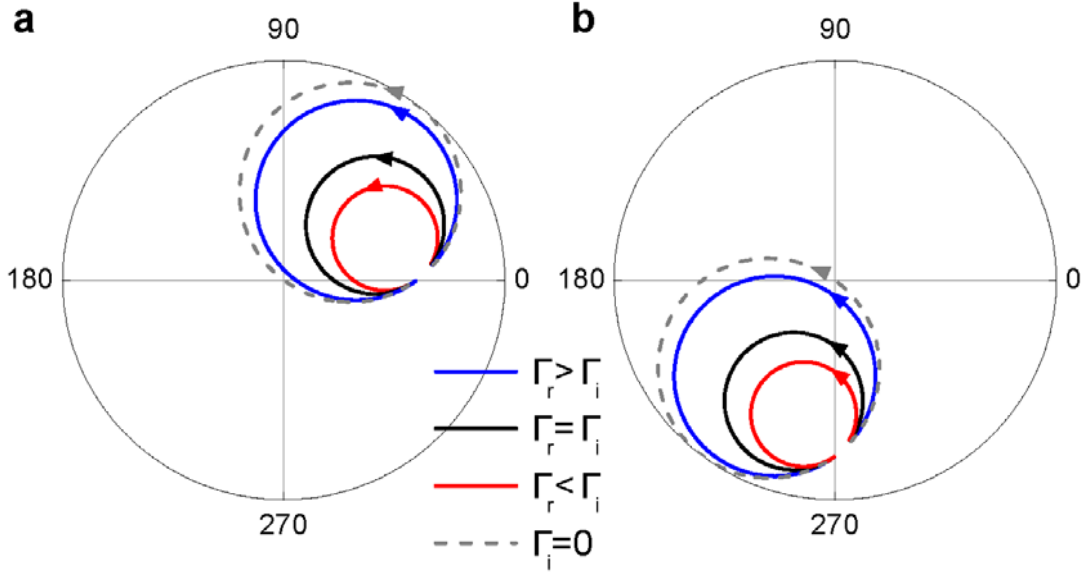

**Figure S14 | CMT analysis of a two-port, single-mode resonator.** Smith curves for **a**, $r$ and **b**, t according to Eq. (S4). Four representative cases are plotted: a) $\Gamma_i = 0.05 f_0, \Gamma_r = 0.1 f_0$ (blue curve); b) $\Gamma_i = 0.1 f_0, \Gamma_r = 0.1 f_0$ (black curve); c) $\Gamma_i = 0.15 f_0, \Gamma_r = 0.1 f_0$ (red curve); d) $\Gamma_i = 0, \Gamma_r = 0.1 f_0$ (gray curve). The reflection and transmission coefficient of the background medium is set as $r_0 = 0.6$ and $t_0 = -0.8i$ without loss of generality.

## VII. Possible ways to further improve the phase modulation range of our metasurfaces

The varying resistivity of graphene is directly responsible for critical transitions in our metasurfaces. Larger variation of graphene resistivity leads to larger range in phase modulation. Given a fixed maximum gate voltage, larger variation in resistivity can be achieved in graphene samples with higher mobility. To demonstrate the feasibility of this idea, we performed FDTD simulations on metasurfaces corresponding to device A and device B in Fig. 3, in which both the graphene and the SU8 spacer layer assume a better, but still realistic, quality (for SU8 this means less intrinsic loss). Our simulation results show that the phase modulation range of the device can indeed be improved to 300° at 0.465 THz (up from the currently observed value of 243°) (Fig. S15). Further increase of phase modulation range may be obtained by improving the doping capability, and by placing the two resonant frequencies closer to each other.

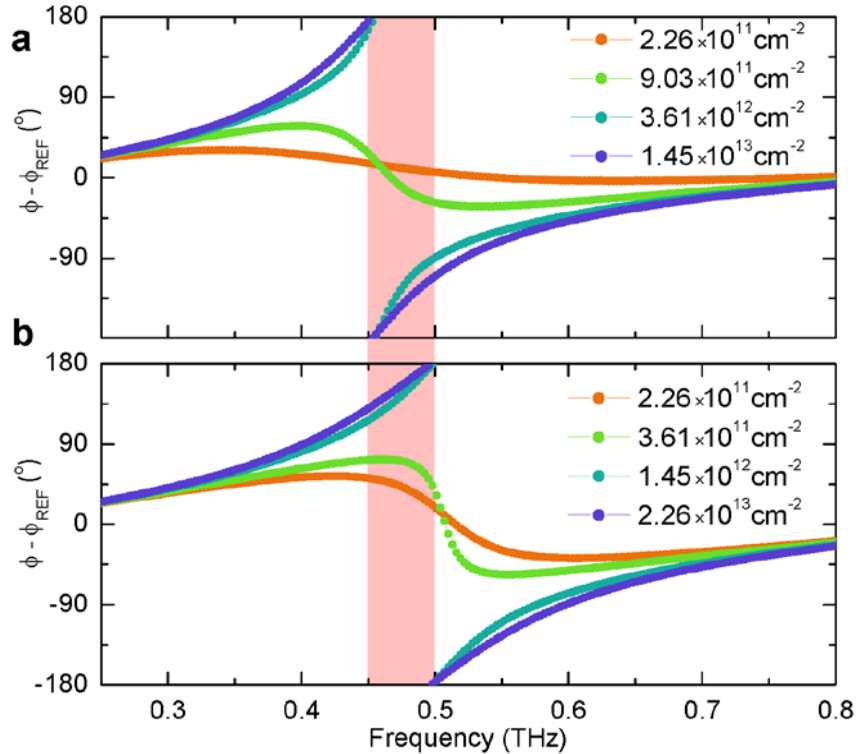

**Figure S15 | FDTD simulations for the phase modulations in two model metasurfaces.** The metasurfaces simulated in **a** and **b** have the same geometry as that of device A and B in Fig. 3, respectively. The graphene and SU8 spacer layer in the simulation, however, assume higher qualities. The electron scattering rate and the residue carrier concentration in graphene are set as 3 THz and $2.26 \times 10^{11} cm^{-2}$, respectively (compared with 4 THz and $\sim 10^{12} cm^{-2}$ in the experiment). Im($\varepsilon$) of SU8 is set as 0.1, compared with 0.28 in reality. The phase modulation range in the shaded region is improved from 243° (in the experiment, Fig. 3) to 300° (this simulation).